\documentclass[twocolumn,nofootinbib,superscriptaddress]{revtex4}	
\usepackage{graphicx}
\usepackage{amsmath}

\newcommand{\SKIP}[1]{}				%	Skip text
\newcommand{\bld}[1]{\mbox{\boldmath $#1$}}	
\newcommand{\smbld}[1]{\mbox{\boldmath\scriptsize $#1$}}	
\newcommand{\beq}{\begin{equation}}
\newcommand{\eeq}{\end{equation}}
\newcommand{\beqar}{\begin{eqnarray}}
\newcommand{\eeqar}{\end{eqnarray}}
\newcommand{\third}{\mbox{${1\over3}$}}         %       1/3
\newcommand{\eps}{\varepsilon}
\newcommand{\bfr}{\boldsymbol{r}}	
	
\newcommand{\rme}{{\rm e}}
\newcommand{\del}{\partial}                     %       partial
\newcommand{\MeV}{{\rm MeV}}			%	MeV
\newcommand{\fm}{{\rm fm}}                      %       fm
 
\newcommand{\HQ}{HQ}

\begin{document}
\title{Non-equilibrium phase transition
	in relativistic nuclear collisions:\\ 
	Importance of the equation of state}
\author{Jan Steinheimer} \email{steinheimer@fias.uni-frankfurt.de}
\affiliation{
Nuclear Science Division, Lawrence Berkeley National Laboratory,
Berkeley, California 94720, USA}
\affiliation{Frankfurt Institute for Advanced Studies,
Johann Wolfgang Goethe Universit\"at,\\
Ruth-Moufang-Stra{\ss}e\ 1, 60438 Frankfurt am Main, Germany}
\author{J{\o}rgen Randrup} \email{JRandrup@LBL.gov}
\affiliation{
Nuclear Science Division, Lawrence Berkeley National Laboratory,
Berkeley, California 94720, USA}
\author{Volker Koch} \email{VKoch@LBL.gov}
\affiliation{
Nuclear Science Division, Lawrence Berkeley National Laboratory,
Berkeley, California 94720, USA}

\date{\today}

\begin{abstract}
Within the context of relativistic nuclear collisions aimed at exploring
hot and baryon-dense matter,
we investigate how the general features of the expansion dynamics, 
as well as a number of specific observables,
depend on the equation of state used in dynamical simulations
of the non-equilibrium confinement phase transition.

\end{abstract}


\maketitle

%========================================================================
\section{Introduction}

One of the central themes of current research in strong-interaction
physics is the exploration and characterization of strongly
interacting matter at high temperatures and/or densities. 
In the laboratory, strongly interacting matter at
high temperature and density is created in collisions of heavy nuclei 
at relativistic energies
and experiments to study these reactions are carried  out at
various accelerator facilities, 
particularly the Relativistic Heavy Ion Collider (RHIC) at BNL
and the Large Hadron Collider (LHC) at CERN. 
In the universe, strongly interacting matter at high density
but very low temperature may exist in the core of neutron stars, 
and its properties control the maximum mass and radius of those stars. 

Experiments at RHIC and LHC have demonstrated that 
collisions at nucleon-nucleon center-of-mass energies of 
$\sqrt{s_{NN}} \geq 60$\,GeV create matter at very high temperature $T$ 
and nearly vanishing net-baryon density, $\rho$,
i.e. $\rho/T^3 \ll 1$. 
Recent continuum-extrapolated lattice QCD calculations 
with physical quark masses
show that such matter exhibits a crossover transformation from an 
interacting hadronic gas to a quark-gluon fluid at a pseudo-critical
temperature of $T_\times\simeq 155$\,MeV
 \cite{Aoki:2006we,Aoki:2006br,Bazavov:2011nk}.

Nuclear collisions at lower energies produce systems 
at finite net-baryon densities
with the corresponding baryon-number chemical potential $\mu$
being of the order of the temperature or higher. 
This region is difficult to explore with lattice QCD techniques 
due to the fermion sign problem 
but for sufficiently small values of $\mu/T$ 
the pressure may be expanded in powers of $\mu/T$,
providing a hint about the finite-$\mu$ behavior
\cite{Allton:2002zi,Kaczmarek:2011zz,Borsanyi:2012cr,%
Borsanyi:2013hza,Gavai:2008zr}. 
Alternative methods use an imaginary chemical potential
\cite{deForcrand:2008vr} or apply re-weighting techniques
\cite{Fodor:2001pe,Fodor:2004nz}.
However, all these approaches are restricted to small values of $\mu$ and
have limited predictive power at the large chemical potentials. 
The theoretical exploration to higher net baryon densities
must therefore rely on models.

Among the many effective models for QCD thermodynamics,
those that couple constituent quarks to mesonic fields and the Polyakov loop 
have become popular because they include chiral-symmetry breaking and
a ''thermal'' confinement of quarks. 
Such approaches are the Polyakov Nambu--Jona-Lasino (PNJL) 
\cite{Fukushima:2003fw,Ratti:2005jh}
and the Polyakov Quark-Meson (PQM) \cite{Schaefer:2007pw} models. 
Recently also results based on functional renormalization group (FRG) 
approaches have been investigated 
\cite{Pawlowski:2005xe,Berges:2000ew,Schaefer:2006sr,Herbst:2013ail}.
These models can be brought into agreement with lattice results
at vanishing net baryon density but their results for the finite-density 
domain depend significantly on the parametrizations used.
A particular weakness in most of these models
is the absence of explicit hadronic degrees of freedom in the confined
phase.
There exist only few examples where the equations of state (EoS) includes 
an approximately realistic description of the hadronic phase. 
These equations of states are constructed either by matching a purely hadronic
model to an MIT-bag EoS or by combining a PNJL/PQM-type model 
with a chiral hadronic model \cite{Steinheimer:2010ib,Turko:2013taa}. 
Obviously, these models need to be constrained both by lattice QCD results 
at zero and small chemical potential and by known properties of nuclear matter
at small temperatures and densities up to about twice nuclear matter density 
\cite{Steinheimer:2011ea}, as well as by astrophysical observations 
on neutron star properties \cite{Dexheimer:2012eu}.

As already discussed in \cite{Hempel:2013tfa} there is a
  qualitative difference between the various
  effective models that do not include explicit hadronic
  degrees of freedom and those which include the hadronic degrees of
  freedom.  Many of the  aforementioned  Polyakov loop models, as well as
  various incarnations of the linear sigma model and Nambu-Jona-Lasinio model
  \cite{Ferroni:2010ct}, belong to the former group. These models
  typically predict coexistence between the (deconfined and/or
  chirally symmetric) dense phase and the vacuum at vanishing
  temperature. Consequently the coexistence pressure at $T=0$
  vanishes and, as a result, it remains small even at moderate
  temperatures. This unrealistic behavior is akin to the well known liquid gas
  phase transition and subsequently we will refer to models in this
  group as LG models. 
  In reality, however, at vanishing temperature
  the (deconfined and/or chirally symmetric) dense phase coexists with
  compressed nuclear matter. Therefore the coexistence pressure is
  finite and large. Indeed, as will be discussed below, it will likely
  be larger than the (pseudo) critical pressure at vanishing net-baryon
  density. This qualitative difference between LG models and
more plausible models will lead to qualitative
  differences for the expansion dynamics of the 
  system, as we shall show.

Experimentally, different regions of the phase diagram of strongly interacting
matter can be explored with nuclear collisions by varying the energy,
thereby influencing the phase trajectory 
along which the bulk of the created system evolves \cite{Arsene:2006vf}. 
However, the successful discovery of a phase transition
requires the measurement of suitable observables
that are sensitive to such a structure in the phase diagram.
Since the identification of signal observables
is a challenging task, due to the complexity of the collision,
one must invoke dynamical simulations.
It is, therefore, essential to develop transport models
that are capable of describing the evolution of the system 
in the presence of a phase transition,
in particular propagate it correctly under the influence of
the associated mechanical instabilities
\cite{PhysRep,Randrup:2003mu,Sasaki:2007db,Randrup:2009gp,RandrupAPH22,
Mishustin:1998eq,Bower:2001fq,Paech:2003fe,Nahrgang:2011mg}.
Such a dynamical model was recently presented 
\cite{Steinheimer:2012gc,Steinheimer:2013gla} and
the present work uses it to explore the sensitivity of various
observables to the presence of a first-order coexistence region
in the phase diagram.

The purpose of the present paper is twofold: First we wish to explore
the qualitative differences between LG-type equations of state and the
more palusible ones that have phase coexistence between compressed
nuclear matter (rather than vacuum)
and a denser (deconfined and/or chirally symmetric) phase. 
In particular we are interested in generic differences in the
fluid dynamic time evolution. To this end we will study two illustrative
equations of state that serve as representatives for the two types of
models, LG and ``realistic'', as discussed above. The second purpose of
the paper is to elucidate the sensitivity of various observables to
the existence of a first-order phase coexistence region and 
its associated instabilities. 

Because we will be working in the
framework of fluid dynamics, the only relevant information is the
equation of state. Therefore, our results will not discriminate
between a phase transition driven by chiral restoration or one driven by
deconfinement. Therefore we will subsequently refer to the (deconfined and/or
chiral symmetric) dense phase simply as the quark phase (QP).
Finally, we will restrict our studies to
situations with only one first-order coexistence region in the phase diagram. 
The potentially interesting case of separate chiral and
deconfinement transition will be left for future work.

The presentation is organized as follows: 
In the next section, we compare an EoS with a hadron-quark-type 
phase transition with the EoS of a PQM model
and discuss their qualitative differences,
with a view towards identifying potential phase-transition observables.
These two equations of state are then used
to carry out various calculations with our dynamical model
which consists of finite-density ideal fluid dynamics augmented by
a gradient term in the local pressure \cite{Steinheimer:2012gc}. Based
on these results, we discuss the suitability of the two models for
semi-realistic tests of the sensitivity of various observables for the
QCD phase structure. Specifically, we discuss the effects of the
non-equilibrium phase transition on composite-particle production
as well as two-particle angular correlations.

Throughout this paper, for notational simplicity,
we denote net baryon density by $\rho$
and the associated baryon-number chemical potential by $\mu$.

%========================================================================
\section{The \HQ\ Equation of State}

To investigate the effects of the instabilities associated with a 
first-order phase transition between a hadronic gas and a quark-gluon plasma,
we need to employ a suitable two-phase EoS
(which we shall refer to as the \HQ\ equation of state).
Although significant progress has been made in understanding the
thermodynamical properties of each of the phases separately,
our current understanding of the phase coexistence region is not yet
on firm ground.  
We therefore follow Ref.\ \cite{Randrup:2010ax}
and employ a conceptually simple approximate
EoS obtained by interpolating between an ideal hadron gas, 
augmented by a $\rho$-dependent interaction energy,
and an idealized quark-gluon plasma. 
As already noted in the Introduction, we view this EoS
 as a representative of those equations of state that exhibit phase
  coexistence between compressed nuclear matter and the dense quark phase.

Furthermore, as a reference, we subsequently define a partner EoS
obtained by eliminating the instabilities by means of Maxwell constructions.

The confined phase is approximated as an ideal gas of pions,
nucleons, and antinucleons, plus an interaction term,
so the total hadronic pressure is of the form
\beq
p^H = p_\pi+p_N+p_{\bar N}+p_w\ .
\eeq
The contributions from the various hadron species are
\beqar
p_\pi(T)\! &=&\! -g_\pi T\int {d^3 k\over(2\pi)^3}
	\ln[1-\rme^{-\epsilon_\pi(k)/T}]\ ,\\
p_N(T,\mu_0)\! &=&\! g_N T\int {d^3 k\over(2\pi)^3}
	\ln[1+\rme^{-(\epsilon_N(k)-\mu_0)/T}]\ ,\\
p_{\bar N}(T,\mu_0)\! &=&\! g_N T\int {d^3 k\over(2\pi)^3}
	\ln[1+\rme^{-(\epsilon_N(k)+\mu_0)/T}]\ ,
\eeqar
where $g_\pi=3$, $g_N=2\times2=4$, and $\epsilon(k)^2=m^2+k^2$
with $m_\pi=140\,\MeV$ and $m_N=940\,\MeV$.
The net baryon density
$\rho^H=\rho_N-\rho_{\bar N}=\del(p_N+p_{\bar{N}})/\del\mu_0$ then follows,
\beq
\rho^H(T,\mu_0) = 
g_N \int_{m_N}^\infty{p\epsilon d\epsilon\over2\pi^2}
{\sinh\mu_0/T\over\cosh\mu_0/T+\cosh\epsilon/T}\ .
\eeq
Finally, the interaction contribution is given by
$p_w(\rho)=\rho\del_\rho w(\rho)-w(\rho)$, where
$w(\rho) = [-A\rho/\rho_s+B(\rho/\rho_s)^2]\rho$
is the interaction energy density.
The strength coefficients $A$ and $B$ have been adjusted
so that nuclear matter saturates at $\rho_s\!=\!0.153\,\fm^{-3}$
and the associated compression modulus is $K=K_N+K_w=300\,{\rm MeV}$; 
the binding energy of nuclear matter is then also roughly reproduced
\cite{Randrup:2010ax}.
The Fermi level $\mu_0$ and the chemical potential $\mu$ are related by
$\mu=\mu_0+\del_\rho w(\rho)/\del\rho$
and the entropy density is then $\sigma^H(T,\mu)=\del p^H(T,\mu)/\del T$.

The deconfined phase was taken as an ideal gas of massless gluons 
and massless {\em ud} quarks with a bag constant $B=300\,\MeV/\fm^3$
\cite{Randrup:2010ax},
\beq
p^Q\ =\ p_g+p_q+p_{\bar{q}}-B\ ,
\eeq
where $p_g=g_g(\pi^2/90)T^4$ with $g_g=2\times8=16$ is the gluon pressure
while the quarks and antiquarks contribute
\beq
p_q+p_{\bar{q}} = g_q
\left[{7\pi^2\over360}T^4+{1\over12}(\third\mu)^2T^2
	+{1\over24\pi^2}(\third\mu)^4\right],
\eeq
with $g_q=2\times3\times2=12$.
The net baryon density in the plasma is then
\beq
\rho^Q\ =\ {\del\over\del\mu} p^Q(T,\mu)\ 
=\ {2\over3}(\third\mu)T^2+{2\over3\pi^2}(\third\mu)^3\ ,
\eeq
while the entropy density is
\beq
\sigma^Q\ =\ {\del\over\del T}p^Q(T,\mu)\ =\ 
{74\over45}\pi^2T^3+2(\third\mu)^2T\ .
\eeq

At zero temperature and zero chemical potential,
the pressure of the nucleon gas vanishes
while that of the quark gas is equal to $-B$.
The confined phase is then the thermodynamically favored one.
However, as the chemical potential is raised,
the plasma pressure increases faster than the hadronic pressure,
so the curves $p^H(T=0,\mu)$ and $p^Q(T=0,\mu)$
cross at a certain value of $\mu$, 
above which the deconfined phase is favored.
This phase crossing procedure can be repeated for any temperature.
As $T$ is increased, the crossing value of $\mu$ decreases steadily,
becoming zero at $T_{\rm max}$,
above which $p^Q$ is larger than $p^H$ for any value of $\mu$.

For the construction of the two-phase equation of state,
it is convenient to work in the  canonical representation.
Then the condition of phase coexistence, 
{\em i.e.}\ same temperature, chemical potential, and pressure 
at two different densities $\rho_1$ and $\rho_2$,
amounts to the condition that $f_T(\rho)$, 
the free energy density at a specified temperature as a function of density,
exhibit a local concavity.
The two coexistence densities are then those at the touching points
of the associated common tangent.
This is readily seen since $\mu_T(\rho)=\del_\rho f_T(\rho)$
implies that the two chemical potentials are then equal, 
$\mu_T(\rho_1)=\mu_T(\rho_2)$,
and since the tangent at $\rho_i$ is given by 
$t_i(\rho)=f(\rho_i)+(\rho-\rho_i)f'(\rho_i)$ 
the fact that $t_1(\rho)=t_2(\rho)$ immediately implies 
that also the pressures are equal, $p_T(\rho_1)=p_T(\rho_2)$.

At zero temperature, the hadronic free energy density, $f_{T=0}^H(\rho)$,
starts at zero but grows more rapidly than the $f_{T=0}^Q(\rho)$,
which starts at $B$.
Therefore the two curves cross and, furthermore,
because they both have positive curvature, a common tangent exists
between two densities $\rho_1$ and $\rho_2$.
This feature signals the occurrence of phase coexistence.
In order to describe the associated transition region,
we follow Ref.\ \cite{Randrup:2010ax} and employ a spline procedure.
Thus, for each temperature below a critical value $T_{\rm crit}<T_{\rm max}$,
we employ a set of matching densities,
$\rho_T^H<\rho_1$ and $\rho_T^Q>\rho_2$,
which define three distinct density regions.
The hadron EoS is used in the low-density region,
$f_T(\rho\leq\rho_T^H)=f_T^H(\rho)$,
the plasma EoS is used in the high-density region,
$f_T(\rho\geq\rho_T^Q)=f_T^Q(\rho)$,
and in the intermediate density region the free energy density
is represented by a spline polynomial $\tilde{f}_T(\rho)$
that matches the values, slopes, and curvatures of $f_T(\rho)$
at $\rho_T^H$ and $\rho_T^Q$.

\begin{figure}[t]	%       -----------------------------------------
\includegraphics[width=0.5\textwidth]{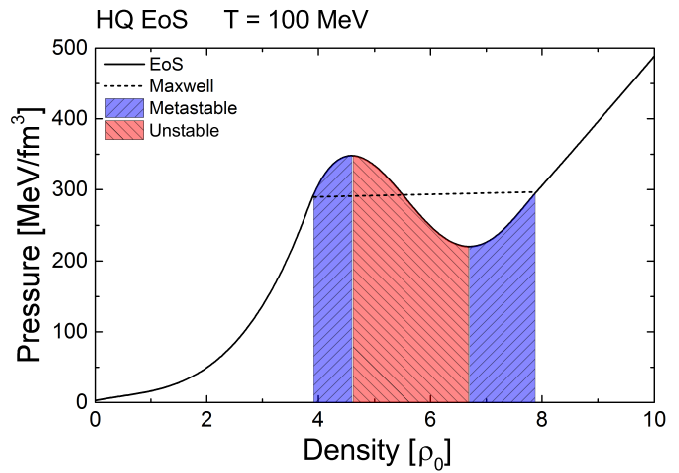}
\caption{(Color online) 
Illustration of the equation of state of the \HQ\ model:
The pressure $p$ as a function of the net baryon density $\rho$
at the fixed temperature $T=100$\,MeV (solid curve)
and the corresponding Maxwell construction (dashed line). 
The middle (red) shaded area shows the mechanically unstable density region
while the adjacent (blue) shaded areas show the meta-stable regions
(in which matter is mechanically stable but thermodynamically unstable).
}\label{1}
\end{figure}

By suitable adjustment of the matching densities,
it is possible to obtain a resulting free energy density curve $f_T(\rho)$
that displays an ever weakening phase transition as $T$ is raised,
up to a critical value, $T_{\rm crit}\approx143$\,\MeV,
above which the phase transformation occurs as a smooth crossover.
The resulting pressure curve, 
$p_T(\rho)=\rho\partial_\rho f_T(\rho)-f_T(\rho)$,
is shown in Fig.\ \ref{1} for $T=100$\,\MeV.

		%       -----------------------------------------

In Ref.\ \cite{Randrup:2010ax} the focus was on subcritical temperatures, 
$T<T_{\rm crit}$, so for each $T$ the spline points were adjusted
so the resulting $f_T(\rho)$ would exhibit a concave anomaly,
ensuring that there would be two densities, $\rho_1(T)$ and $\rho_2(T)$,
for which the tangent of $f_T(\rho)$ would be common,
signaling phase coexistence.
Ref.\ \cite{Steinheimer:2012gc} extended the EoS to $T>T_{\rm crit}$
by using convex splines, as is characteristic of single-phase systems.
Having constructed $f_T(\rho)$ for a sufficient range of $T$ and $\rho$,
we may obtain the energy density,
$\eps_T(\rho)=f_T(\rho)-T\partial_T f_T(\rho)$, by suitable interpolation 
and then tabulate the EoS, $p_0(\eps,\rho)$, 
on a convenient Cartesian lattice in the mechanical densities 
$\eps$ and $\rho$.

%========================================================================
\section{The PQM Model}

As an alternative, we shall also consider a Polyakov Quark Meson model 
 as a representative of a class of  models based on 
the coupling of constituent quarks to mesonic fields and the Polyakov loop. 
Other approaches of this type include the PNJL and FRG models 
in different realizations \cite{Fukushima:2003fw,Megias:2004hj,Ratti:2005jh,%
Roessner:2006xn,Sasaki:2006ww,Ghosh:2006qh,Schaefer:2007pw,Fukushima:2008wg}.
These models have become quite popular because they include 
the correct degrees of freedom at high temperatures 
(namely weakly interacting quarks and gluons) 
and they have an effective thermodynamic description 
of the confinement transition and the chiral symmetry breaking
 that is in reasonable agreement with lattice results at $\mu=0$. 
Within chiral fluid dynamics, this model has been recently applied
in relativistic nuclear collisions for the purpose of
investigating the effects of a first-order deconfinement transition 
and the critical endpoint \cite{Herold:2013bi,Herold:2013qda}.
 However, these models yield a rather poor description of the
 confined phase, as they allow for only (uncorrelated) three-quark states. 
Therefore, they typically severely underestimate the pressure
in the hadronic phase, as we shall discuss subsequently.
In addition, at zero temperature this type of model as well as
  sigma and NJL models \cite{Ferroni:2010ct} typically predict
  phase coexistence between the vacuum and the dense quark phase,
  similar to the well known liquid gas transition. 
  We thus view the
  PQM model as a generic representative of LG-type models.

 The effective thermodynamic potential of the PQM model can be written as
 \begin{equation} \Omega\ =\
	-\frac{T}{V}\ln{Z}\ =\ {\cal U}+U_\sigma+\Omega_{q\overline{q}}\ .
 \end{equation}
The meson potential is given by
 \begin{equation} U_{\sigma}\ =\
	\mbox{$1\over4$}\lambda(\sigma^2-\nu^2)^2-c\sigma-U_0
         +g_\omega \lambda_{\omega}^2 \omega^2\ ,
 \end{equation}
where the last term represents a repulsive interaction 
 transmitted by the $\omega$ vector field. 
 The Polyakov-loop potential used in the present work
 is the logarithmic form introduced in Ref.\ \cite{Roessner:2006xn},
 \begin{eqnarray}
         {\cal U}&=&-\mbox{$1\over2$}a(T)\Phi\Phi^*\\ \nonumber
         &+&b(T)\ln[1-6\Phi\Phi^*+4(\Phi^3+\Phi^{*3})-3(\Phi\Phi^*)^2]\ ,
 \end{eqnarray}
 with $a(T)=a_0 T^4+a_1 T_0 T^3+a_2 T_0^2 T^2$, $b(T)=b_3 T_0^3 T$.
 The parameters $a_0=3.51, a_1=-2.47, a_2=15.2$, and $b_3=-1.75$ are fixed as in 
 Ref.\ \cite{Roessner:2006xn}, and $T_0=210$\ MeV.

 A coupling of the quarks to the Polyakov loop is introduced 
 in the thermal energy of the quarks,
 
 \begin{eqnarray}
 \Omega_{q \overline{q}}& = &2 N_f \int \frac{d^3 k}{(2 \pi)^3}\nonumber 
 \left\{ T \ln\left[1+3\Phi\,\rme^{-(\epsilon(k)-\mu_q^*)/T}\right.\right. \\ 
 &~&+\left.3\Phi^* \rme^{-2(\epsilon(k)-\mu_q^*)/T}
	+\rme^{-3(\epsilon(k)-\mu_q^*)/T}\right] \\ \nonumber  
 &+&T \ln\left[1+3\Phi^*\rme^{-(\epsilon(k)+\mu_q^*)/T} \right.\nonumber 
\\ \nonumber 
 &~&+\left.3\Phi\,\rme^{-2(\epsilon(k)+\mu_q^*)/T}\left.
	+\rme^{-3(\epsilon(k)+\mu_q^*)/T}\right]\right\}\ ,
 \end{eqnarray}
 where only the light-quark flavors ($u$ and $d$) are included.
 The thermal heat bath of the quarks contains contributions 
 from one-, two-, and three-quark states,
  of which only the latter do not couple to the Polyakov loop. 
 The quark energy, $\epsilon(k)=\sqrt{k^2+m^{*2}}$, 
includes the effective mass, 
 $m_i^*=m_0+g_{\sigma} \sigma$ with $m_0=6$\,MeV, 
 and the repulsive vector coupling leads to an effective chemical potential, 
 $\mu_q^*=\mu_q-g_{\omega}\omega$ 
that depends on the strength of the quark vector coupling $g_\omega$.
The PQM model parameters used in this work are $\lambda=19.7$,
$\nu=87.6\, \MeV$, $c=1.77 \cdot 10^6 \, \rm{MeV}^3$,
$\lambda_\omega=500 \, \MeV$, $U_0 = 165\cdot 10^6 \, \rm{MeV}^4$,
$g_\sigma = 3.2$, and $g_\omega = 0$ or $g_\omega = 2$. 

 For vanishing vector interaction strength, $g_{\omega}=0$, 
 the PQM model exhibits the desired phase structure,
 namely a smooth crossover at zero chemical potential 
 as well as a first-order phase transition at zero temperature. 
 The critical endpoint is located at $\mu_{\rm crit}\approx480$\,MeV 
 and $T_{\rm crit}\approx165$\,MeV \cite{Schaefer:2007pw}. 

 \begin{figure}[t]	%       -----------------------------------------
 \includegraphics[width=0.5\textwidth]{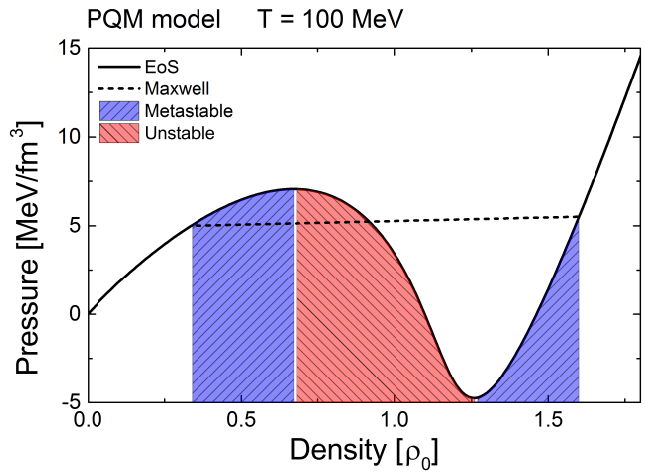}
 \caption{(Color online) 
 Illustration of the equation of state of the PQM model:
 The pressure $p$ as a function of the net baryon density $\rho$
 at the fixed temperature $T=100$\,MeV (solid curve)
 and the corresponding Maxwell construction (dashed line). 
 The middle (red) shaded area shows the mechanically unstable density region
 while the adjacent (blue) shaded areas show the meta-stable regions
 (in which matter is mechanically stable but thermodynamically unstable).
 }\label{2}
 \end{figure}		%       -----------------------------------------

 Using the PQM model, we can calculate a two-phase EoS, 
with an unstable region, by finding the local maxima (unstable states) 
and minima (meta-stable states) of the grand canonical potential.
The resulting EoS is illustrated in Fig.\ \ref{2}.
 It should be noted that the phase transition in the PQM model occurs 
 at density and pressure values that are significantly lower than those
 of the \HQ\ construction discussed above.

 \begin{figure}[t]	%       -----------------------------------------
 \includegraphics[width=0.5\textwidth]{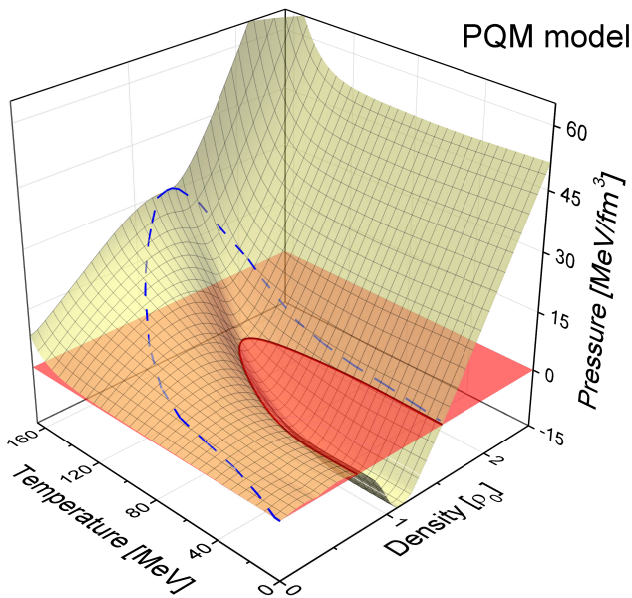}
 \caption{(Color online) The PQM equation of state $p(T,\rho)$:
The pressure a function of both temperature and net baryon density. 
The phase-coexistence boundary is delineated (dashed path)
and the plane of zero pressure is shown (red),
making it easier to see where the pressure is negative.
 }\label{3}
 \end{figure}		%       -----------------------------------------

 A more global impression of the PQM EoS (for $g_{\omega}=0$)
 can be obtained from Fig.\ \ref{3} which shows $p(T,\rho)$,
 the dependence of the pressure on both temperature and baryon density. 
 The pressure is negative in a large part of the phase coexistence region.
 It is also evident that at $T=0$ there is coexistence between 
 the deconfined quark phase and the vacuum,
 a feature that will admit stable quark droplets in the vacuum. 
 However, we know that the true ground state of matter, 
 coexisting with the vacuum,
 is nuclear matter at saturation density and not quark matter. 
 It will become apparent in the following that this inconsistency 
 has strong implications
 for the dynamical evolution of the hot and dense matter
 produced in relativistic nuclear collisions.

It is instructive to consider the temperature dependence of the 
{\em pseudo-critical} pressure, $p_{\rm pc}$, 
the pressure at which the phase transformation occurs.
At subcritical temperatures, $T<T_{\rm crit}$, 
$p_{\rm pc}$ is obviously equal to the coexistence pressure,
but in the cross-over region ({\em i.e.}\ above the critical temperature)
the definition of $p_{\rm pc}$ is not unique
(though different definitions yield qualitatively similar results).
Essential differences between models are then brought out
by the behavior of the transition pressure with temperature, $p_{\rm pc}(T)$.

\begin{figure}[t]	%       -----------------------------------------
 \includegraphics[width=0.5\textwidth]{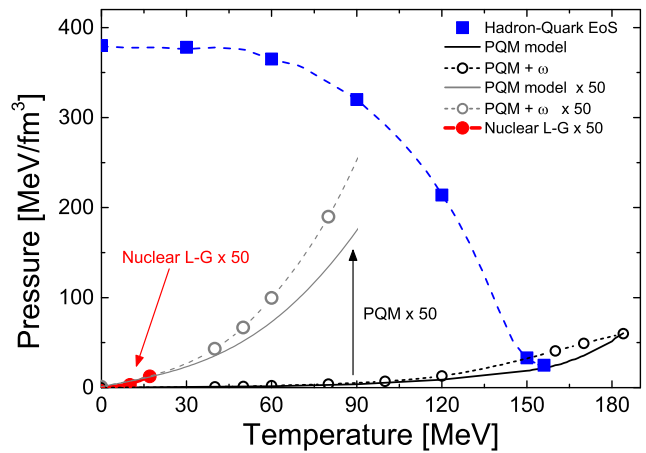}
 \caption{(Color online) 
 The pseudo-critical pressure as a function of the temperature,
$p_{\rm pc}(T)$, 
for both the \HQ\ (blue dashed curve with squares)
 and the PQM model (black curves) models. 
 The black dashed line with open circles shows the result of the PQM model 
 with finite quark vector coupling.
On a scale magnified by a factor of fifty,
the PQM results (grey) are compared with the nuclear liquid-gas transition
(which displays the familiar liquid-gas phase transition).
 }\label{4}
 \end{figure}

Figure \ref{4} shows the temperature dependent pseudo-critical pressure
for the two models considered here.
For a specified temperature $T$,
$p_{\rm pc}^{\rm \HQ}$ is obtained by increasing the chemical potential $\mu$ 
until the ideal hadron-gas and the quark-gluon bag have the same pressure,
while $p_{\rm pc}^{\rm PQM}$ is taken as the pressure along 
the inflection point of the chiral condensate. 

While the resulting pseudo-critical pressures are comparable
at the highest temperatures (corresponding to $\mu\approx0$), 
and consistent with lattice results \cite{Borsanyi:2010cj}, 
they deviate strongly at large chemical potentials:
As the temperature is reduced,
$p_{\rm pc}^{\rm \HQ}$ increases steadily,
as was already noted in previous studies employing models that describe 
the hadron-quark transition \cite{Randrup:2010ax,Hempel:2013tfa}.
By contrast, $p_{\rm pc}^{\rm PQM}$ decreases steadily and vanishes at $T=0$,
a behavior that is a robust feature of the PQM models and 
persists also in the presence of a repulsive quark interaction \cite{Lourenco:2012dx,Pinto:2012aq}.

For comparison, Fig.~\ref{4} also shows the coexistence pressure
for the liquid-gas phase transition in ordinary nuclear matter.
It is quantitatively very similar to the low-temperature part
of the corresponding PQM result.
In particular, these pressures start from {\em zero} at $T=0$
and then {\em increase} steadily with $T$.
This generic liquid-gas behavior differs qualitatively from the \HQ\ result,
for which the pressure {\em decreases} steadily with temperature.

 		%       -----------------------------------------

This analysis suggests that the nature of the PQM phase transition
resembles that of the nuclear liquid-gas transition 
and differs qualitatively from the quark-hadron transition.
For the liquid-gas phase transition,
the dense phase (the liquid) has fewer active degrees of freedom than the
dilute phase (the vapor),
so the liquid-to-gas transition increases the entropy per baryon.
By contrast, for the confinement transition the entropy per baryon
decreases because the dense system (the quark-gluon plasma) has more
active degrees of freedom than the coexisting dilute system (the hadron gas). 
Indeed, from the Clausius-Clapeyron relation it follows that
a decrease in the entropy per baryon from the dense to the dilute phase 
requires that the pressure decrease with temperature \cite{Hempel:2013tfa},
as exhibited in \HQ\ EoS.   
It is important to keep this qualitative difference in mind. 

While at present it is not known which class the QCD phase-transition
belongs to, we note that recent lattice QCD calculations \cite{Borsanyi:2012cr}
have extracted the pressure at the crossover temperature 
$T_\times=153\,\rm MeV$ and vanishing chemical potential 
to be $p(T_\times,\mu=0)\simeq50\,\rm MeV/\fm^3$. 
This corresponds to the pressure of cold nuclear matter at
a density of roughly three times the ground state density, 
$\rho=3 \rho_s$ \cite{Prakash:1988md,prakash}. 
Therefore, unless the hadron-quark transition happens at
densities of $\rho \simeq 3 \rho_s$ or below, the pressure is expected
to rise as we lower the temperature and increase the density,
just as $p_{\rm pc}^{\rm \HQ}(T)$ in Fig.~\ref{4}.

A further inspection of Fig.~\ref{4} shows that even though the
\HQ\ and PQM pseudo-critical pressures are numerically rather similar
in the high-temperature region where $\mu\approx0$,
the slopes of $p_{\rm pc}^{\rm \HQ}(T)$ and $p_{\rm pc}^{\rm PQM}(T)$
are opposite.
Because this is the phase region that can be accessed by lattice QCD methods
it would be interesting to investigate whether 
those can help to distinguish between the models. 
We therefore briefly discuss the prospects for this.

The introduction of the pseudo-critical pressure $p_{\rm pc}(T)$ 
makes it possible to define the associated 
pseudo-critical chemical potential $\mu_{\rm pc}(T)$
determined by $p_{\rm pc}(T)=p(T,\mu_{\rm pc}(T))$.
In lattice QCD the inverse relationship, $T_{\rm pc}(\mu)$,
has been obtained to second order in $\mu$ 
for various definitions of the pseudo-critical line 
\cite{Endrodi:2011gv,Borsanyi:2012cr,Kaczmarek:2011zz},
 \begin{equation}
 T_{\rm pc}(\mu)=T_\times\left[1-\kappa\,\frac{\mu^2}{T_\times^2}\right]\ ,
 \label{eq:pseudo_critical}
 \end{equation}
where we recall that $T_\times=T_{\rm pc}(0)$ 
is the crossover temperature at $\mu=0$.
The curvature of the pseudo-critical line $T_{\rm pc}(\mu)$
depends on the definition of pseudo criticality.
For example, if the inflection point of the chiral condensate is used
then $T_{\times}=153\,\MeV$ and $\kappa = 0.0066$ \cite{Endrodi:2011gv}.
On the other hand, if the inflection point of $s/T^3$ is used
(where $s$ is the entropy density)
then $\kappa = 0.016$, more than a factor of two larger,
and $T_{\times} = 158\,\MeV$, 
using the lattice EoS parametrization from Ref.~\cite{Borsanyi:2012cr}.

In order to determine the slope of the pseudo-critical pressure,
we use that $p(T,\mu>0)$ can be obtained by a Taylor expansion around $\mu=0$
\cite{Allton:2002zi,Gavai:2008zr},
 \begin{equation}
p(T,\mu)\ =\ p(T,\mu=0)
 +T^4\sum_{n>0}{\frac{\chi_{2n}(T)}{(2n)!}\left(\frac{\mu}{T}\right)^{2n}} .
 \end{equation} 
The Taylor coefficients $\chi_n(T)$ are the $n$'th-order 
baryon-number susceptibilities evaluated at vanishing $\mu$, 
 \begin{eqnarray}
 \chi_n(T)\ =\ \frac{\partial^n}{\partial  \left( \mu/T
   \right)^n} \left.\frac{p}{T^4}\right|_{\mu=0}\ . 
 \end{eqnarray}
Thus, for sufficiently small values of $\mu_{\rm pc}$,
the pseudo-critical pressure, $p_{\rm pc}(T)=p(T,\mu_{\rm pc}(T))$, 
is given by 
 \begin{equation}
p_{\rm pc}(T) = p(T,\mu=0) +T^4\sum_{n>0}
{\frac{\chi_{2n}(T)}{(2n)!}\left(\frac{\mu_{\rm pc}(T)}{T}\right)^{2n}} .
 \end{equation}
To second order in $\mu_{\rm pc}$ this can be evaluated using
 Eq.~(\ref{eq:pseudo_critical}). 
In particular, the slope of the pseudo-critical pressure at $\mu=0$ is given by
 \begin{equation}
{\partial\over\partial T}\,p_{\rm pc}(T,\mu=0)|_{T=T_\times} =\
	s(T_\times,\mu=0)-{T_\times^3\over2\kappa}\,\chi_2(T_{\times})\ .
 \end{equation}
The entropy density, $s$,
the curvature of the pseudo-critical line, $\kappa$,
and the second-order susceptibility,  $\chi_2$,
have been calculated on in lattice QCD with physical quark masses
and in the continuum limit \cite{Borsanyi:2012cr,Borsanyi:2011sw}.
Using those values, we then find that the sign of the above slope 
depends on the definition of the pseudo criticality. 
If it is defined by the inflection points of either 
the chiral condensate or the strangeness susceptibility 
we find the slope of the pressure along the pseudo-critical to be
 $\partial_T \,p_{\rm pc} =-2.3 T_\times^3 $ 
or $\partial_T \,p_{\rm pc} =-1.1 T_\times^3$,  respectively. 
In both case the slope is negative meaning the pressure
 increases as we decrease the temperature,
 similar to the \HQ\ result depicted in Fig.~\ref{4}.
If, on the other hand, the pseudo-critical line is defined by 
the inflection point of $s/T^3$, the slope of the pseudo-critical pressure
turns out to be positive, $\partial_T \,p_{\rm pc} = + 2.4 T_\times^3 $. 
This is a simple consequence of the fact that the curvature of 
the pseudo-critical line, $\kappa$,
is a more than a factor of two larger in this case. 

Thus, from the above considerations we must, regrettably, conclude 
that at present lattice QCD calculations cannot distinguish 
between the two scenarios at hand: a liquid-gas type behavior 
where the pseudo-critical pressure increases with temperature 
or a decreasing behavior
as is suggested by phenomenology and implemented in the \HQ\ EoS
used in the present study.

%========================================================================
\section{Fluid Dynamical Clumping}\label{clumping}

Next we employ the two equations of state in dynamical calculations
and look for qualitative and quantitative differences between the results.
Specifically, we employ ideal fluid dynamics to study systems 
as they expand through the unstable phase region. 
Dissipative effects are not expected to play a
decisive role for the spinodal clumping \cite{Randrup:2010ax}, because 
even though the inclusion of viscosity generally tends to slow the growth,
the dissipative mechanisms also lead to heat conduction 
which has the opposite effect and also enlarges the unstable region.
The equations of motion derived from conservation of four-momentum 
and net baryon number
are solved by means of the code SHASTA \cite{Rischke:1995ir} in which the 
propagations in the three spatial dimensions are carried out consecutively.

In order to obtain a proper description of the spinodal growth rates
in the mechanically unstable region of the phase diagram,
it is essential to introduce finite-density effects
\cite{Randrup:2009gp,Randrup:2010ax}.
This was done in Ref.\ \cite{Steinheimer:2012gc} by augmenting
the pressure obtained from the EoS, $p_0(\eps,\rho)$, by a gradient term,
$\delta p(\bld{r})=-a^2\rho(\bld{r})\nabla^2\rho(\bld{r})\eps_s/\rho_s^2$,
so the local pressure is given by
$p(\bld{r})=p_0(\eps(\bld{r}),\rho(\bld{r}))+\delta p(\bld{r})$.
For the \HQ\ EoS,
the resulting simulation tool was verified to yield the correct growth rates
of the mechanical instabilities in the spinodal phase region
 \cite{Steinheimer:2012gc}.
Furthermore, it was found that the associated spinodal instabilities 
may cause significant amplification of initial density irregularities 
in relativistic lead-lead collisions 
\cite{Steinheimer:2012gc,Steinheimer:2013gla}.

We use this tool to investigate the effect of the different equations of state
on observables for the QCD phase transition at large baryon densities.
To facilitate the comparisons,
we employ the same range $a$ in the gradient term also for the PQM EoS
(namely $a=0.033\,\fm$) which then yields a significantly smaller surface
tension (namely $\approx\!2\,\MeV/\fm^2$ and compared to 
$\approx\!10\,\MeV/\fm^2$ for the \HQ\ EoS).
This is not surprising because the PQM densities tend to be smaller,
as then also the differences between the coexistence densities are,
and the PQM EoS looks rather similar to the LG EoS
which has a smaller surface tension (namely $\approx\!1\,\MeV/\fm^2$).

In particular, we will study the evolution of spherically symmetric systems, 
where the energy and baryon densities ($\epsilon$ and $\rho$) are initialized 
according to a Woods-Saxon distribution,
\begin{equation}\label{WSe}
	\epsilon(\bfr)=\epsilon_{I} [1+e^{(r-c)/w}]^{-1}\ ,
\end{equation}
\begin{equation}\label{WS}
	\rho(\bfr)=\rho_{I} [1+e^{(r-c)/w}]^{-1}\ ,
\end{equation}
where $r$ is the distance to the center 
and $\epsilon_I$ and $\rho_I$ are the corresponding central density values. 
For the radius we use $c=4\,\fm$
and the width parameter is $w=0.5\,\fm$.
To provide initial seeds of fluctuations to be amplified, 
we modify the density profile by adding random fluctuations 
$\delta\epsilon(\bfr)$ and $\delta\rho(\bfr)$ to the local densities, 
$\epsilon(\bfr)$ and $\rho(\bfr)$. 
The amplitudes of the fluctuations are constrained to a maximum of $\pm20\%$ 
of the corresponding local densities (\ref{WSe}) 
and they are uniformly distributed within this range. 
 
For both equations of state, we initialize the system at a temperature
of $100$ MeV and a density that corresponds to a phase point $(\eps,\rho)$
slightly above the unstable region (for the PQM EoS) or just inside
the unstable region (for the \HQ\ EoS). We obtain the strongest
'clumping' when the system is initialized inside the spinodal region. 
For the PQM model such a system would have negative pressure, 
as shown in Fig.\ \ref{3}. 
Consequently, the system will grow denser
until its phase point leaves the unstable region 
and the pressure turns positive. 
Because this initial compression delays the eventual dynamical expansion, 
the PQM calculation is initialized already at a density 
above the unstable region, thus eliminating the delay 
that would otherwise be caused by the negative pressure.

As the system evolves, we extract several quantities that might be sensitive
to the developing instabilities. 
To establish a suitable reference, we also let the systems evolve from the same
initial state but with the corresponding stable EoS
obtained by eliminating the phase-transition instabilities
by means of Maxwell constructions,
thus preserving the features outside the phase-coexistence region.

\begin{figure}[t]	%       -----------------------------------------
\includegraphics[width=0.5\textwidth]{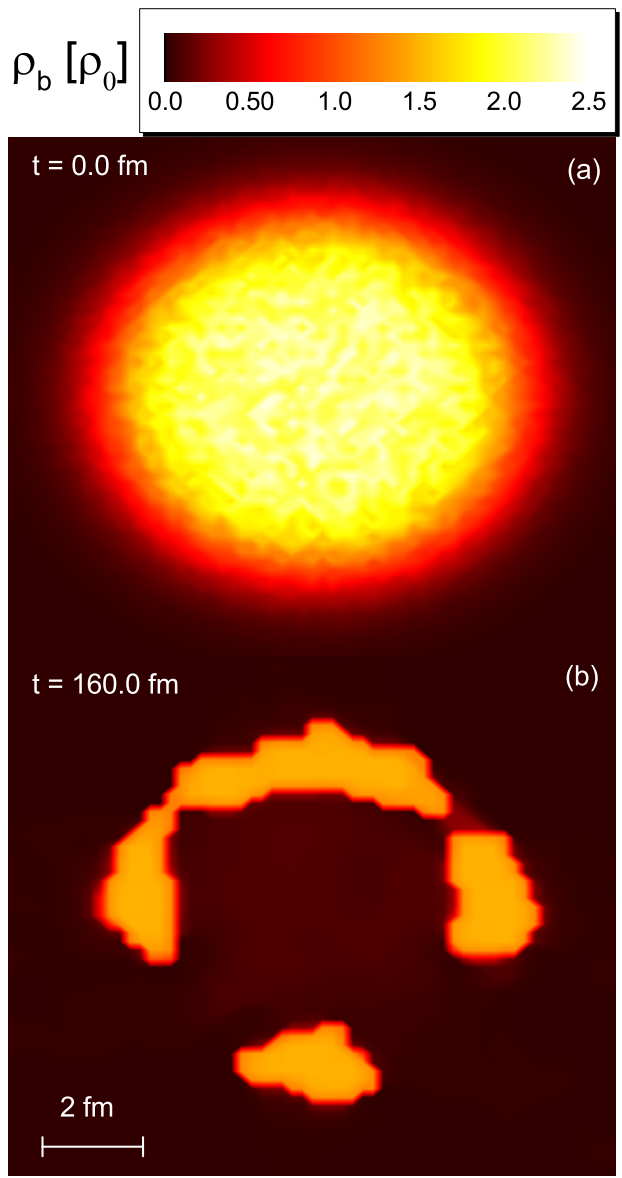}
\caption{(Color online) 
(a): Initial net baryon density distribution in the x-y plane. 
The initial state is constructed to lie just above the unstable region 
of the PQM equation of state.
(b): Net baryon density distribution in the x-y plane after $160\,\fm/c$ 
evolution with the PQM equation of state, that has an unstable region. 
The clusters of high baryon density are clearly visible.
}\label{PQM}
\end{figure}

For the PQM EoS, 
Fig.\ \ref{PQM}(a) shows the initial density distribution in the $xy$ plane, 
as obtained by augmenting Eqs.\ (\ref{WSe}-\ref{WS}) by fluctuations.
This initial state is then evolved with ideal fluid dynamics
(with the gradient term included) using the unstable EoS.
The resulting density distribution after $160\,\fm/c$ 
is shown in Fig.\ \ref{PQM}(b). 
Even after such a long time one finds meta-stable clusters 
of high baryon density with an asymmetric angular distribution,
an observation that was first reported in \cite{Herold:2013qda}.
This phenomenon is a direct consequence of the liquid-gas nature 
of the PQM model, which predicts phase coexistence between the dense phase 
and the vacuum at vanishing temperature. 
As a consequence, the pressure even at $T=100\,\MeV$ 
is very small (see Fig.\ \ref{2}) resulting in a slow expansion. 
In addition, due to the small pressure difference between 
the dense phase and the vacuum, the clusters generated during
the unstable stage are nearly stable and therefore live for a long time,
before they eventually slowly disappear on a timescale of hundreds of $\fm/c$. 

		%       -----------------------------------------

\begin{figure}[t]	%       -----------------------------------------
\includegraphics[width=0.5\textwidth]{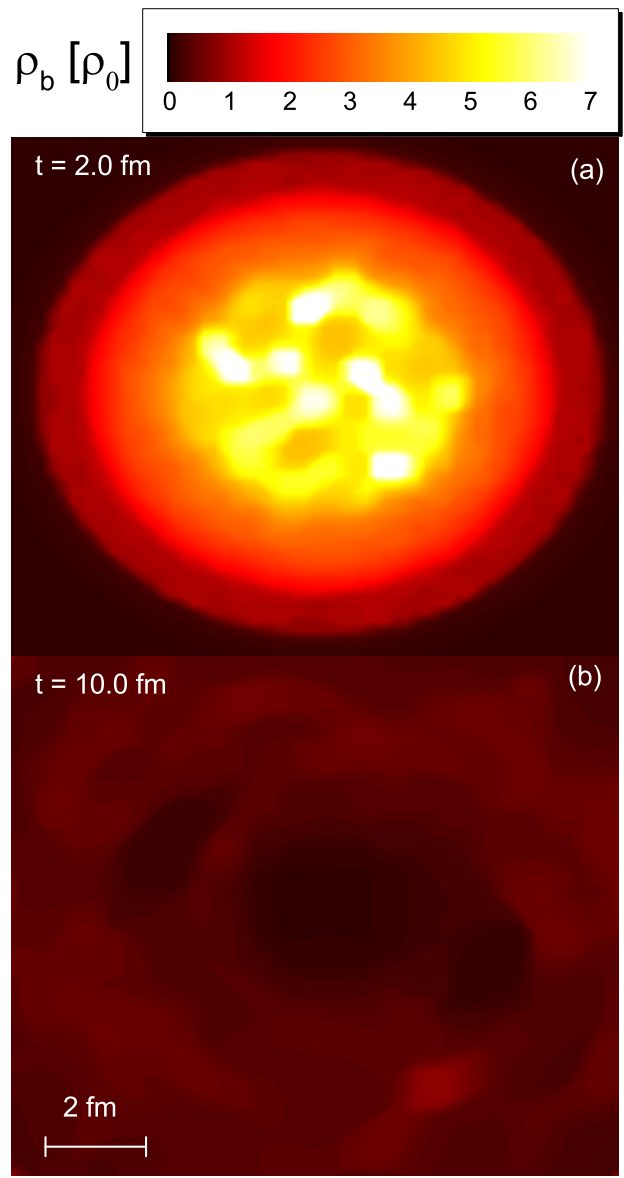}
\caption{(Color online) 
The net baryon density in the $xy$ plane,
$\rho(x,y,0,t)$, after $t=2\,\fm/c$ $(a)$ and $t=10\,\fm/c$ $(b)$,
as obtained with the unstable \HQ\ EoS:
Clumps of dense quark matter are formed at early times $(a)$
but they disappear after further evolution $(b)$.
}\label{HQ}
\end{figure}		%       -----------------------------------------

\begin{figure}[t]	%       -----------------------------------------
\includegraphics[width=0.5\textwidth]{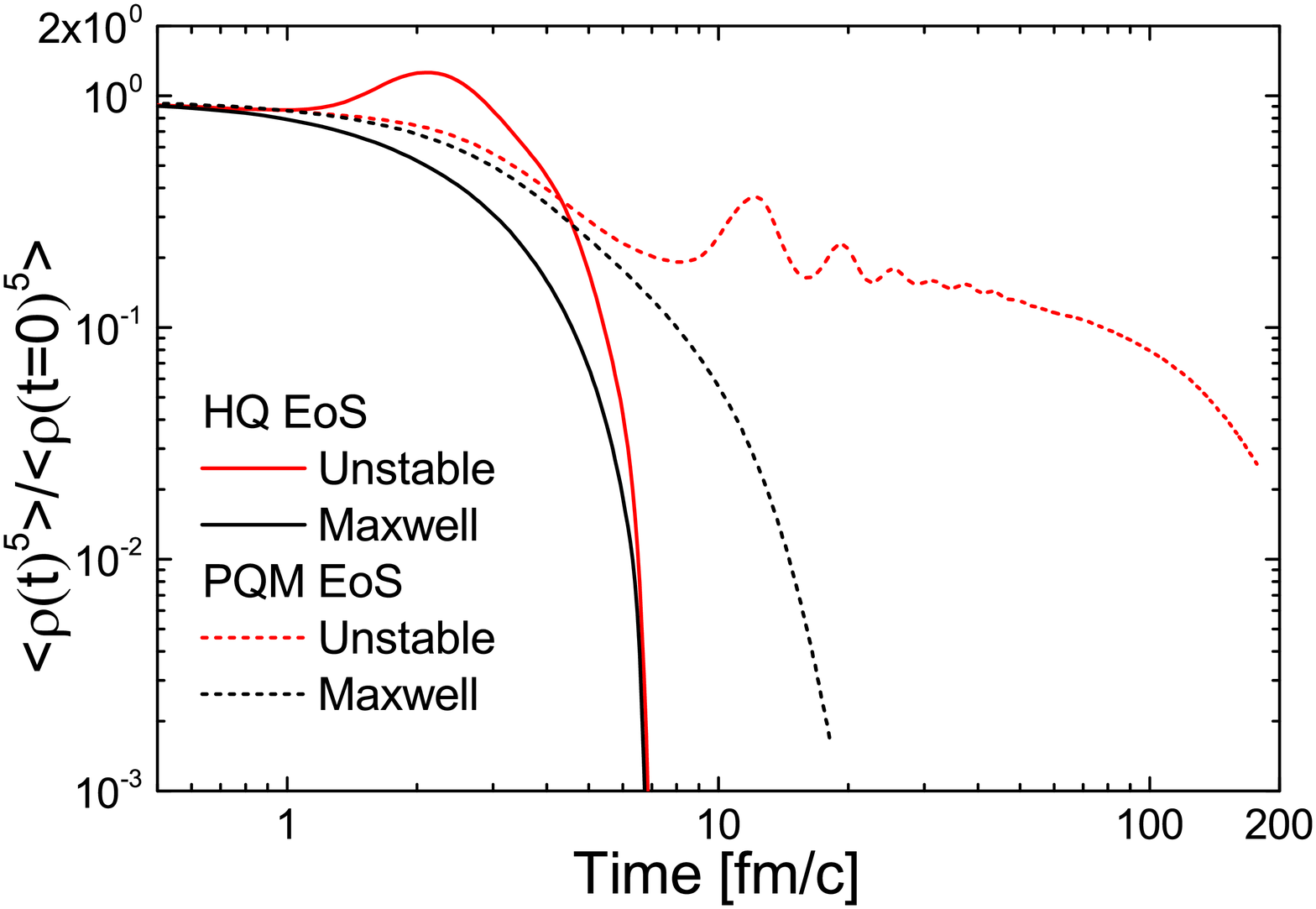}
\caption{(Color online) The fifth density moment (see Eq.\ (\ref{m})), 
normalized to its initial value, 
as a function of time for the \HQ\ (solid) and the PQM (dashed) EoS. 
In each case, the upper curve (red) depicts results obtained with the
unstable EoS, while the lower curve (black) has been obtained 
with the corresponding Maxwell partner EoS.
}\label{7}
\end{figure}		%       -----------------------------------------

The situation is quite different in case of the \HQ\ EoS.
As in the calculation based on the PQM EoS,
the fluctuations are quickly amplified to form small clusters,
illustrated in Fig.\ \ref{HQ}(a). 
However, in contrast to the PQM, the QGP clusters disappear rather quickly 
after the surrounding hadronic matter expands and dilutes. 
This is due to the very high pressure  difference between the dense phase 
and the vacuum, as depicted in Fig.~\ref{1}. 
The diluted density distribution, after a time of only $10\,\fm/c$, 
is shown in Fig.\ \ref{HQ}(b), 
where it is evident that the previously formed dense clusters 
have been completely washed out.
 
Consequently, the two scenarios, PQM and \HQ,
exhibit the same qualitative features 
but lead to quantitatively different time scales and amplitudes. 
It is therefore useful to now estimate the quantitative differences 
in observables resulting from the two equations of state.

%========================================================================
\section{Observables}

In this section we investigate the sensitivity of various proposed 
observables to the formation of clusters of deconfined matter.

%------------------------------------------------------------------------
\subsection{Powers of the density}

The density moments provide a convenient global quantification
the irregularities in the evolving net baryon density $\rho(\boldsymbol{r},t)$
\cite{Steinheimer:2012gc},
\begin{equation}\label{m}
\langle\rho(t)^N\rangle	\equiv\		{1\over A}
\int\rho(\boldsymbol{r},t)^N\rho(\boldsymbol{r},t)\,d^3\boldsymbol{r}\ ,
\end{equation}
where $A=\int\rho(\boldsymbol{r},t)d^3\boldsymbol{r}$ is the (constant)
net baryon number of the system.
As already shown in Ref.\ \cite{Steinheimer:2012gc}, 
the higher moments are more sensitive to the magnitude of the fluctuations
and we show results for $N=5$ here.

Figure \ref{7} shows the time evolution of the fifth density moment,
divided by its initial value $\langle{\rho(t=0)}^5\rangle$,
for the two different equations of state considered.

For the \HQ\ EoS the instabilities in the 
coexistence region lead to local density enhancements
which manifest themselves in the bump of the density moment.
Relative to the evolution with the Maxwell partner EoS,
the maximum enhancement of the fifth density moment is about a factor of 3-4.
But once the surrounding medium has become sufficiently dilute
these clumps again dissolve resulting in a rapid drop of the density moment. 
The largest degree of enhancement occurs
when most of the system is inside the unstable phase region. 
After the system leaves this region the density quickly becomes similar 
to that obtained with the stable partner EoS.

The PQM EoS yields a quite different picture. 
Because of the low pressure at the phase-coexistence line, 
the dynamical evolution is considerably slower and
the density moment starts to increase only after a long time.
And because the created clumps are almost stable,
and only decay very slowly when embedded in a very dilute gas, 
the density enhancements become much larger 
than was the case for the \HQ\ EoS. 
This qualitative difference is clearly brought out
in  Fig.\ \ref{7} for the fifth density moment.
We also note its oscillatory behavior 
which can be ascribed to the longevity of the created clusters:
As the clusters are formed, 
the inflowing baryon current will over-compress the dense quark droplet
which in turn will cause it to subsequently expand slowly 
bringing it again slightly inside the unstable region,
thus causing a second (somewhat weaker) over-compression and so on. 

The calculation also shows that the PQM density moment starts
to decrease appreciably only after a very long time ($100-200\,\fm$).
A similar slow decrease of the PQM density moments
was also observed in a different simulation \cite{Herold:2013qda}.
This clearly shows how the liquid-gas type properties of the deconfinement 
transition implied by the PQM EoS may yield potentially misleading results.

%------------------------------------------------------------------------
\subsection{Composite production}

Another potential signal observable is the production yield of 
composite particles ({\em i.e.}\ light nuclei such as deuterons and tritons) 
from phase-space coalescence of nucleons.  
In the simple coalescence picture,
the phase-space density of a composite with baryon number $A$ 
is proportional to the $A$'th power of the one-particle 
nucleon phase-space density 
\cite{Csernai:1986qf,Dover:1989zu,Sato:1981ez,Dover:1991zn,Scheibl:1998tk}.
Therefore, because the spinodal instabilities significantly enhance higher
powers of the density, as demonstrated in  Fig.~\ref{7}, one might
naively expect that the composite production should be enhanced. 
However, as the local baryon density is being enhanced
also the local excitation energy per baryon is increased
and that will decrease in the coalescence yield,
almost compensating the increase due to the higher density, 
as we shall demonstrate below.

First let us recall that the coalescence model gives the
production yield in terms of the nucleon yield as
\cite{Csernai:1986qf,Dover:1989zu,Sato:1981ez,Dover:1991zn,Scheibl:1998tk},
\begin{equation}
	E_A\,\frac{dN_A}{d^3\bld{P}_A}\ \propto\ F_A\,
\left(\left[E\frac{dn}{d^3\bld{p}}\right]_{\smbld{p}=\smbld{P}_A/A}\right)^A .
\end{equation}
Here $F_A$ is a coalescence factor  related to the quantum mechanical 
overlap integral between the wave functions of the parent nucleons
and the composite to be formed; it depends only weakly on $A$. 
Given the simple assumptions made in this work,
{\em e.g.}\ for the initial conditions, 
a calculation of the absolute yield of composite nuclei is not warranted.
Instead we focus on the relative composite yields
in the presence or absence of spinodal instabilities.

\begin{figure}[t]	%       -----------------------------------------
\includegraphics[width=0.5\textwidth]{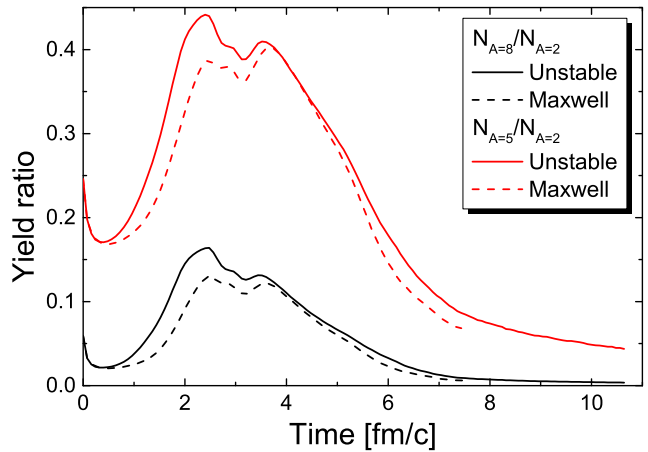}
\caption{(Color online) 
The ratios of the $A=5$ and $A=8$ nuclei over the deuteron yield 
as functions of time. For simplicity, the degeneracy factors 
for $A=8$, $A=5$, and $A=2$ have been set to equal values.
Shown are results for the \HQ\ equation of state with an unstable region 
(solid lines) and with a Maxwell construction (dashed lines).  
}\label{8}
\end{figure}

This simplifies our calculation considerably, 
because Ref.~\cite{Csernai:1986qf} showed that for a non-degenerate
system the coalescence yield and the thermal production yield 
are identical up to factors involving the coalescence radius 
and the binding energy of the composites. 
Because these factors will be the same in all scenarios we are considering,
we should get a good estimate for the relative enhancement 
by simply integrating the local thermal composite populations.
To this end, we apply the Cooper-Frye prescription 
on an isochronous hypersurface, 
using values of the temperature and chemical potential 
of a hadron gas corresponding to the local energy and baryon densities.
The total population of light nuclei with mass number $A$ is then given by
\begin{equation} 
	N_A= \int d^3\bld{p}\,d^3\bld{r} \ f_{A}(\bld{r},\bld{p})\ ,
\end{equation}
with $f_{A}(\bld{r},\bld{p})\propto
	\exp[-(\sqrt{m_A^2+p^2}-\mu_A(\bld{r}))/T(\bld{r})]$,
where we have used an effective degeneracy of one
because we are interested only in ratios.
Furthermore, we assume that $\mu_A(\bld{r})=A\mu(\bld{r})$
and $m_A=Am_N$, where $m_N$ is the nucleon mass. 
Figure \ref{8} shows the number of composites with baryon number 
$A=5$ and $A=8$ relative to the deuteron yield ($A=2$). 
It is obvious that the relative composite populations
increase at the time when the baryon density is enhanced, 
{\em i.e.}\ when the moments of the baryon number distributions increase. 
However, the increase in the population of composite nuclei 
is much smaller than the increase in the density moments.
Indeed, we observe a  maximum increase in the relative composite production 
of only up to $20\%$, 
as compared to what is obtained with the corresponding Maxwell EoS. 

\subsection{Angular correlations}

As can be seen from Fig.~\ref{PQM}, 
there seems to be strong irregularities in the angular distribution 
of the baryon density in the case of strong clumping. 
It has been asserted in \cite{Herold:2013qda} that this will lead to large 
higher moments of the spacial angular distribution of the baryon number.
In this section we will investigate the angular irregularities further 
by calculating the two-particle angular correlations of baryons, 
in both coordinate and momentum space. 

In position space 
these can be computed directly from the fluid dynamical simulations,
\begin{equation}
	V_n^{\rm{pos}}\ =\ {1\over N^2}\sum_{i,j=1}^N
	\rho_i\,\rho_j\,\cos(n(\phi_i^{\rm pos}-\phi_j^{\rm pos}))\ ,
\end{equation}
where $\rho_i$ is the net baryon density 
at the computational lattice point $i$,
$\phi_i^{\rm pos}$ is its azimuthal angle, and $N=\sum_i\rho_i$.

However, experiments usually measure the momentum-space asymmetries $v_n$ 
and momentum correlations rather than coordinate-space asymmetries. 
The azimuthal momentum distribution of the emitted
particles is commonly expressed as
\begin{equation}
	\frac{dN}{d \Phi^{\rm mom}} \propto 
	1 + \sum_{n=1}^{\infty}{2 v_n \cos{(n(\Phi^{\rm mom}-\Psi_n))}}
\end{equation}
where $v_n$ is the magnitude of the $n$'th
order harmonic term relative to the angle of the initial-state
spatial plane of symmetry $\Psi_n$.
To calculate the two-baryon correlations in momentum space
we need the relative angle between the momenta of baryons
coming from the cells $i$ and $j$, $\Phi_{ij}^{\rm mom}$,
\begin{equation}\label{vn}
	V_n^{\rm mom}\ =\ \frac{1}{N_{ij}}\sum_{ij}
	\rho_i\,\rho_j\,\cos(n\Delta\Phi_{ij}^{\rm mom})\ .
\end{equation}
The angle of the momentum of a baryon coming from cell $i$ 
is determined by sampling the Cooper-Frye equation 
\cite{Cooper:1974mv,Petersen:2008dd} with the local values 
of four velocity $u_{\nu}$, temperature $T$, and chemical potential $\mu$. 
This method will create a statistical uncertainty 
which can be minimized by making multiple samplings of $V_n^{\rm{mom}}$ 
and averaging Eq.\ (\ref{vn}) over all samplings.

\begin{figure}[t]	%       -----------------------------------------
\includegraphics[width=0.5\textwidth]{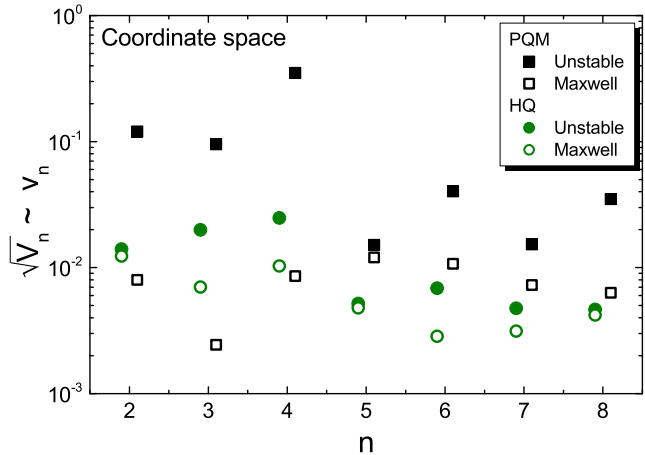}
\caption{(Color online) The position-space two-baryon angular correlation 
strength $V_n^{\rm{pos}}$ for various values of $n$
as obtained with the \HQ\ (circles) and PQM (squares) EoS.
The results obtained with an unstable EoS (solid) 
are contrasted with those employing the Maxwell partner (open).}
\label{9}
\end{figure}

\begin{figure}[t]	%       -----------------------------------------
\includegraphics[width=0.5\textwidth]{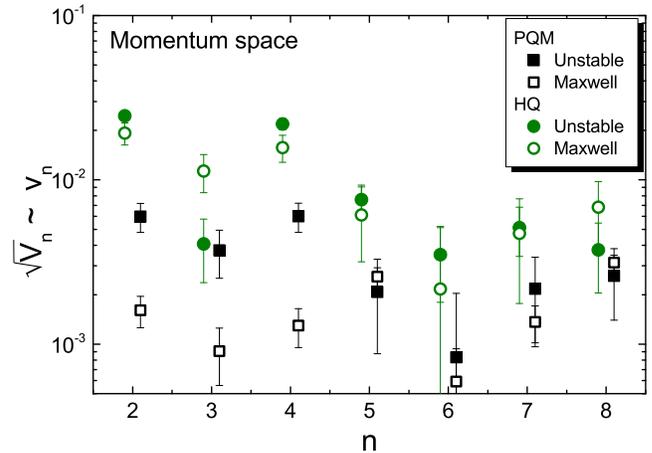}
\caption{(Color online) The momentum-space two-baryon angular correlation 
strength $V_n^{\rm{mom}}$ for various values of $n$
as obtained with the \HQ\ (circles) and PQM (squares) EoS.
The results obtained with an unstable EoS (solid) 
are contrasted with those employing the Maxwell partner (open). 
}\label{10}
\end{figure}

It can be shown \cite{Luzum:2010sp,Aamodt:2011by} 
that the extracted correlation 
measures \mdseries $V_n^{\rm{mom}}$ are related to the squares of the Fourier 
coefficients of the angular distributions:
\begin{equation}
	V_n^{\rm{mom}}= v_n^2
\end{equation}
assuming no corrections due to momentum conservation and resonance decays.
Figures \ref{9} and \ref{10} show our results 
for the coordinate and momentum space angular correlations. 
We compare results with the PQM model, at the end of the evolution
(see figure \ref{7}), and the \HQ\ model, also at the end of the evolution.

As expected from Ref.\ \cite{Herold:2013qda}, 
the coordinate-space correlations from the PQM model are very large
when the system goes through the unstable region. 
As shown in section \ref{clumping},
using the unstable PQM equation of state will create large quark clusters,
containing most of the system's baryon number. 
This of course leads to strong correlations in coordinate-space
and several moments are even larger than $10 \%$.
When the Maxwell constructed equation of state is used 
no clusters are formed and the baryon
matter is distributed more homogeneously in coordinate-space, 
resulting in no strong spacial correlations. 

The calculation with the \HQ\ model shows
much smaller coordinate-space correlations, of at most $2-3 \%$. 
This result is also understandable, as the produced clusters contain
only a small fraction of the system's baryon number and quickly 
become unstable and expand. 
However, the clumping obtained with the unstable EoS 
is still larger than with the corresponding Maxwell construction
which again showed no density enhancement.

Figure \ref{10} shows how these coordinate-space irregularities 
are transformed into momentum-space correlations. 
The high pressure along the coexistence line in the \HQ\ model leads to an 
efficient translation of coordinate space irregularities into momentum space. 
Importantly, the difference between the unstable EoS and the Maxwell EoS
is not very large. Only a factor $1.4$ increase in the momentum correlation
amplitude is observed. The considerable increase in the baryon density,  
due to the clumping in our simulations with the unstable EoS, does not lead to 
a considerable change in the pressure gradients. 
Consequently, the additional spacial anisotropies 
are not reflected strongly in the final momentum anisotropies.

Due to the small pressure in the PQM equation of state, the coordinate-space 
asymmetries here cause even less momentum-space asymmetries, which is 
already seen for the Maxwell constructed EoS that has no instabilities. 
Even after an evolution time of $160$ fm, 
which is much longer than the expected lifetime of the fireball, 
the unstable PQM EoS shows no strong angular momentum-space correlations, 
even though it had very large angular correlations in coordinate space.

\section{Summary}
  
We have investigated the effect of qualitatively
different types of two-phase equation of state 
on the expansion dynamics and
on possible experimental signals of the expected QCD phase transition 
at large baryon densities. 
We considered two equations of state representing  two
  qualitatively different EoS classes. On the one
  hand, as an example for an EoS exhibiting phase coexistence between
  compressed nuclear matter and the dense quark phase, we considered an
  EoS that is constructed by interpolating between
	a hadron gas and a quark-gluon phase. 
	On the other hand, as a representative for the class of
  liquid-gas type equations of state, we used the EoS from a PQM-like
  model.

We find that the PQM model shows a transition 
that is very similar to that of the liquid-gas transition in nuclear matter
and thus this model differs qualitatively from the HQ model with regard to
the thermodynamic properties near the phase coexistence line. 
We also compared the effective equations of state with a Taylor
expansion of recent lattice QCD data. Unfortunately, the predicted
slope of the pressure along the pseudo-critical line at vanishing chemical
potential depends of the specific definition of pseudo criticality.
Therefore, at present lattice results 
are not able to discriminate between the liquid-gas
type behavior of the PQM equation of state and the \HQ\ equation of state.   
This is unfortunate, as the two types of EoS yield dramatically different
pressures in the high-$\mu$ region.

The qualitative differences between the two equations of state
examined in this work lead to significant quantitative differences 
in the time evolution of fireballs that expand through the 
respective unstable region of the phase diagram. 
In the PQM model the lifetime of quark clusters is orders of magnitude longer
 than what is usually expected for the timescales of heavy-ion collisions 
and it predicts stable dense quark matter droplets at zero temperature,
i.e.\ coexisting with the vacuum.

The instabilities associated with the presence of a first-order phase 
transition lead to large irregularities in the spatial distribution 
of the baryon number.
However, these irregularities are not translated into significant 
momentum correlations.
We also show that quark cluster formation only leads to an small increase 
in the production of composite nuclei, of less than $20\%$. 
This increase is  only observed when the system 
is inside the unstable region. 
Once the system has left the coexistence region
and dispersed all signals seem washed out.

In conclusion, we have shown that the qualitative differences between
the PQM and \HQ\ equations of state
lead to considerable differences in the dynamical 
evolution of the system. 
Given the arguments provided in this paper, we believe that these
  qualitative differences persist independent of the specific EoS model
  adopted of either the LG type or a more plausible type. 
However, the observation of these differences is a challenging task.
The two rather intuitive observables explored here have shown little
sensitivity to either the instabilities associated with the
first-order phase co-existence or the dynamical evolution of the system. 
Thus the key question on how to detect those instabilities, if indeed present, 
remains unsettled.

Furthermore, the qualitative behavior of the pressure along the
pseudo-critical line in the QCD phase plane is not yet known.
Lattice QCD calculations of higher-order Taylor coefficients 
for the pseudo-critical line and the associated pressure
might help to clarify this crucial issue. 
Meanwhile, as long as it remains an open question
whether the pseudo-critical pressure increases or decreases,
both possibilities must be considered when validating
any potential phase-transition signal with dynamical calculations 
similar to the ones presented here.

~\\
\section{Acknowledgments}
The authors would like to thank M. Prakash for discussions regarding the
equation of state of neutron stars. 
VK would like to the thank the EMMI Rapid Reaction Task Force on
``Probing the Phase Structure of Strongly Interacting Matter with
Fluctuations'' where some of the ideas for this work  were initiated.
This work was supported by GSI and Hessian initiative for excellence (LOEWE) 
through the Helmholtz International Center for FAIR (HIC for FAIR) 
and by the Office of Nuclear Physics
in the U.S.\ Department of Energy's Office of Science
under Contract No.\ DE-AC02-05CH11231.
JS was supported in part by the Alexander von Humboldt Foundation
as a Feodor Lynen Fellow.

%%%%%%%%%%%%%%%%%%%%%%%%%%%%%%%%%%%%%%%%%%%%%%%%%%%%%%%%%%%%%%%%%%%%%%%%%%%%%%

\end{document}